\documentclass[journal]{IEEEtran}
\ifCLASSINFOpdf
\else
   \usepackage[dvips]{graphicx}
\fi
\usepackage{url}

\usepackage{stackengine}
\def\delequal{\mathrel{\ensurestackMath{\stackon[1pt]{=}{\scriptstyle\Delta}}}}
\usepackage[table,xcdraw]{xcolor} 
\usepackage{graphicx}
\usepackage{caption}
\usepackage{amsmath}
\usepackage{xcolor}
\usepackage{cite}
\usepackage{flushend}

\definecolor{amber}{rgb}{1.0, 0.75, 0.0}
\definecolor{almond}{rgb}{0.94, 0.87, 0.8}
\definecolor{blond}{rgb}{0.98, 0.94, 0.75}
\definecolor{cornflowerblue}{rgb}{0.39, 0.58, 0.93}
\definecolor{lavenderblue}{rgb}{0.8, 0.8, 1.0}
\definecolor{lightskyblue}{rgb}{0.53, 0.81, 0.98}
\definecolor{lime(web)(x11green)}{rgb}{0.0, 1.0, 0.0}
\definecolor{lime(colorwheel)}{rgb}{0.75, 1.0, 0.0}
\definecolor{persianpink}{rgb}{0.97, 0.5, 0.75}
\definecolor{mistyrose}{rgb}{1.0, 0.89, 0.88}
\definecolor{ticklemepink}{rgb}{0.99, 0.54, 0.67}
\definecolor{salmonpink}{rgb}{1.0, 0.57, 0.64}
\definecolor{richbrilliantlavender}{rgb}{0.95, 0.65, 1.0}
\definecolor{pink}{rgb}{1.0, 0.75, 0.8}

\usepackage{color, colortbl}
\definecolor{Gray}{gray}{0.9}
\definecolor{LightCyan}{rgb}{0.88,1,1}
\usepackage[first=0,last=9]{lcg}

\begin{document}

\title{Space-Time Block Coded Reconfigurable Intelligent Surface-Based Received Spatial Modulation}
\author{Ferhat Bayar, Onur Salan, Haci Ilhan, Senior \IEEEmembership{Member, IEEE},  and Erdogan Aydin 
 \thanks{F. Bayar and O. Salan are with the Scientific and Technological Research Council of Turkey (TUBITAK), Kocaeli, Turkey (e-mails: ferhat.bayar@tubitak.gov.tr, onur.salan@tubitak.gov.tr).}
 \thanks{H. Ilhan is with the Department of Electronics and Communications Engineering, Yildiz Technical University, Istanbul, 34220, Turkey, e-mail: ilhanh@yildiz.edu.tr (Corresponding author: Hacı Ilhan).}
 \thanks{E. Aydin is with the Department of Electrical and Electronics Engineering, Istanbul Medeniyet University, Istanbul 34857, Turkey (e-mail: erdogan.aydin@medeniyet.edu.tr)}
}

\maketitle
\begin{abstract}
Reconfigurable intelligent surface (RIS) structures reflect the incident signals by adjusting phase adaptively according to the channel condition where doing transmission in order to increase signal quality at the receiver. Besides, the spatial modulation (SM) technique is a possible candidate for future energy-efficient wireless communications due to providing better throughput, low-cost implementation and good error performance.  Also, Alamouti's space-time block coding (ASBC) is an important space and time coding technique in terms of diversity gain and simplified ML detection. In this paper, we proposed the RIS assisted received spatial modulation (RSM) scheme with ASBC,  namely RIS-RSM-ASBC. The termed RIS is portioned by two parts in the proposed system model. Each one is utilized as an access point (AP) to transmit its Alamouti coded information while reflecting passive signals to the selected received antenna.   The optimal maximum likelihood (ML) detector is designed for the proposed RIS-RSM-ASBC scheme. Extensive computer simulations are conducted to corroborate theoretical derivations. Results show that RIS-RSM-ASBC system is highly reliable and provides data rate enhancement in contrast to conventional RIS assisted transmit SM (RIS-TSM), RIS assisted transmit quadrature SM (RIS-TQSM),  RIS assisted received SM (RIS-RSM), RIS assisted transmit space shift keying with ASBC (RIS-TSSK-ASBC) and RIS-TSSK-VBLAST schemes.
\end{abstract}
\IEEEpeerreviewmaketitle
\section{Introduction}
According to Cisco, all mobile traffic worldwide will reach 77 exabytes per month by the end of 2022. In this context, 5G, which provides enhanced mobile broadband, ultra-reliable and low-latency communications and massive machine-type communications were completed in June 2018. While mmWave and orthogonal frequency-division multiplexing (OFDM) efforts increase to meet 5G standard targets, it has been determined that there is no single technology that can support 5G application requirements \cite{basar2019wireless}. Therefore, the researchers have started studies about new technologies such as non-orthogonal multiple access (NOMA), multiple-input and multiple-output (MIMO), terahertz communication \cite{chen2019survey}, which will form the basis of 6G technology. The key performance metrics of 6G aims to meet are spectrum and energy efficiency,  ultra-reliable low-latency communication, coverage and mobility, as well as providing intelligent and dynamic systems that quickly adapt to changing environmental conditions and application types \cite{giordani2020toward}. In this context, schemes such as media-based modulation \cite{seifi2016media, naresh2016media,eaydin_MBM}, spatial modulation (SM) \cite{aydin2015novel} and space shift keying (SSK) \cite{jeganathan2009space} use reconfigurable antenna to get rich scattering environments. SM and SSK techniques have been presented with other techniques due to index modulation's (IM) spectral and energy efficiency. The NOMA-based SM scheme provides increased spectral efficiency without additional complexity. Generalized code index modulation (GCIM) aided SM  (GCIM-SM) scheme is proposed to increase the communication system's energy efficiency and data rate while reducing energy consumption \cite{GCIM_fatih}. Also, in the literature, there are working about the cooperative protocol \cite{aydin2020c} and physical layer security system \cite{jiang2017secrecy} which work with IM techniques to improve overall system error performance. 

The researchers focused on the controlling propagation environment to improve the quality of service (QoS) using intelligent reflecting surfaces. The reconfigurable intelligent surface (RIS) scheme uses a large number of small, low-cost and passive reflecting elements. The RIS adjusts the phase of the incident signal and then reflects it to the destination without any energy source \cite{liaskos2018new}. In \cite{basar2020reconfigurable}, the researcher proposed RIS aided SSK (RIS-SSK) and RIS based SM (RIS-SM) schemes in collaboration with IM systems on the RIS scheme. Also, \cite{basar2020reconfigurable} and \cite{salan2021performance} provide the error performance analysis of RIS-based IM over the Rayleigh fading channel and Weibull fading channel, respectively. Alamouti scheme was used with a RIS scheme in \cite{khaleel2020reconfigurable}. The RIS-based Alamouti scheme provides improved error performance compared to the classical Alamouti scheme. \cite{li2021space} proposes RIS based Alamouti scheme which enables the RIS to transmit coded information while reflecting the incident SSK signals. Two new deep neural networks (DNN) aided cooperative RIS techniques are proposed for cooperative communication systems in \cite{sagir2022deeplearning}. In one model of this study, DNN is embedded in the RIS unit; in the other model, DNN is deployed at the destination for symbol detection instead of a maximum likelihood (ML) detector. RIS-assisted media-based modulation (MBM) scheme is introduced in \cite{Onal2021}.

\begin{figure*}[t]
\centerline{\includegraphics[width=18cm,height=8cm]{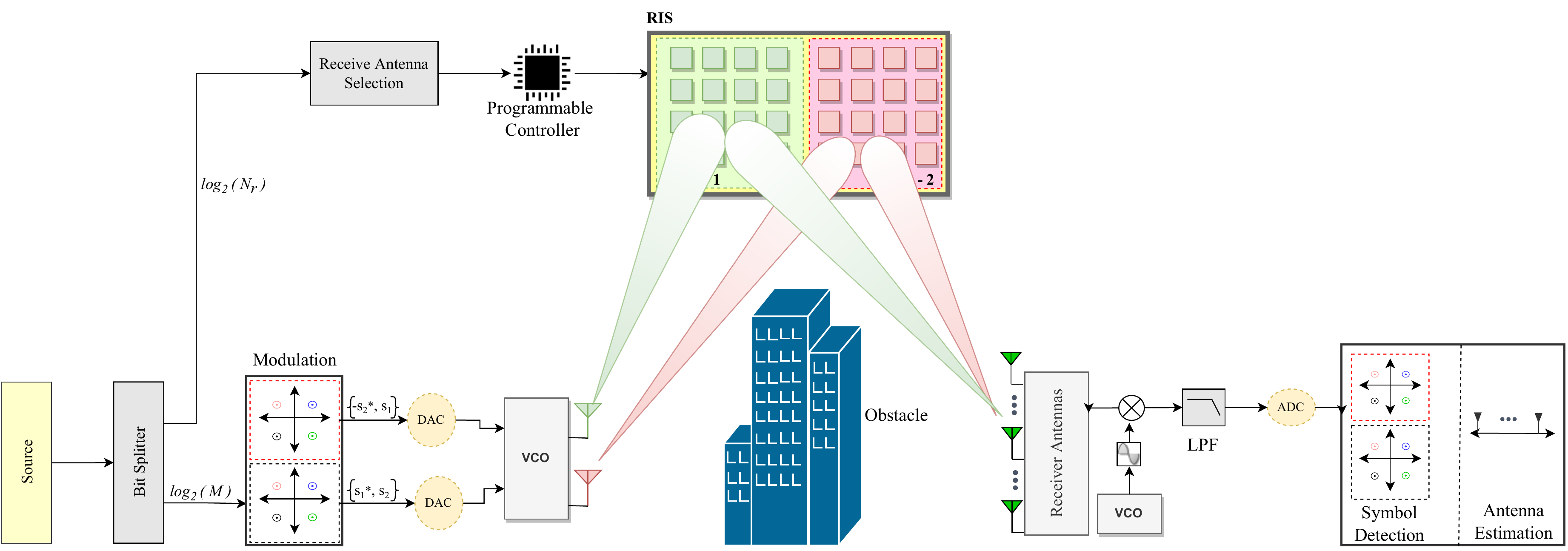}}
\caption{System model of the RIS-RSM-ASBC scheme.}
\centering\label{Fig:SystemModel}
\end{figure*}

In this paper, we propose a RIS-based received SM scheme with Alamouti space-time block coding to improve the error performance of conventional RIS-based systems. Also, the theoretical derivations of error performance are presented for the proposed scheme.

\subsection{Motivation and Contribution}
Motivated by the provides ultra-reliable communication, especially at low SNR and achieving better error performance compared to the RIS-RSM \cite{basar2020reconfigurable} in the literature, we investigate the RIS-RSM-ASBC in this paper. Also, we study RIS-RSM-ASBC for the potential data rate improvement using the SM scheme at the receiver side instead of the transmitter.
The novel contribution of our paper can be summarized as follows:
\begin{itemize}
    \item For the first time in literature, we propose an Alamouti's coded RIS-assisted SM at the receiver using optimal ML Detector.
    \item We represent the theoretical error performance of the proposed scheme and provide useful insights.
    \item We compare the error performance of the proposed scheme and other schemes such as conventional RIS-TSM, RIS-TQSM, RIS-RSM, RIS-TSSK-ASBC, and RIS-TSSK-VBLAST.
    \item It is aimed to get over 10 dB better error performance in the RIS-RSM-ASBC compared to the RIS-TSSK-ASBC system.
    \item Also, another goal is to achieve better error performance with less reflectors in the proposed scheme compared to RIS-RSM.
    \item We show that the proposed scheme provides performance enhancement and enables highly reliable transmission, especially at low SNR.
    \item Lastly, the simulations and theoretical results are evaluated at low SNR to show the proposed scheme enables highly reliable transmission.
\end{itemize}

\subsection{Notations}
Matrices and vectors are shown in boldface uppercase and boldface lowercase letter, respectively. $\left(\cdot\right)^*$, $\left(\cdot\right)^T$, and $\left(\cdot\right)^H$ define complex conjugation, transposition, and Hermitian transposition, respectively. $[\cdot]^{-1}$ and $\mathbf{I}_{n\times n}$ stand the inverse of a matrix and the identity matrix with $n\times n$, respectively. $M$ defines modulation order. $X_{\Re}$ and $X_{\Im}$ denote the real and imaginary part of $X$, respectively. Parameters/symbols used in this paper are also listed in the Table \ref{TableListOfSymbols}. 
\subsection{Paper Organization}
The remainder of this paper is organized as follows. Section II introduces the principle of the Alamouti's coded RIS-based Received SM scheme and defines the received signal model. Section III presents the derivation of the proposed system's BEP. The analytical and Monte Carlo simulation results and complexity analysis are given in Section IV. Finally, Section V concludes the paper.

\definecolor{pastelpink}{rgb}{1, 0.82, 0.86}
\definecolor{wildwatermelon}{rgb}{1, 0.65, 0.8}
\begin{table}
  \caption{List of Parameters/Symbols}
\centering
\scalebox{0.99}{%
\begin{tabular}{l c c}
		\rowcolor{wildwatermelon}
\hline\hline 
Parameters/Symbols  & Definition\\\hline\hline \rowcolor{pastelpink}
$\mathrm{E}\left[X\right](\mu_X)$  & Mean of RV \textit{X}
\\ 
 \rowcolor{pastelpink}
$\mathrm{Var}\left[X\right](\sigma^2_{X})$  & Variance of RV \textit{X} 
\\ \rowcolor{pastelpink}
$\mathrm{E}[X^n](m_{X(n)})$ &  $n^{th}$ moments of RV \textit{X}
\\ \rowcolor{pastelpink}
$f_X\left(\cdot\right)$ & Probability density function (PDF) of RV \textit{X}
\\ \rowcolor{pastelpink}

$M_X\left(\cdot\right)$  & Moment generating function (MGF) of RV \textit{X}
\\ \rowcolor{pastelpink}
$\Gamma(\cdot)$ & Gamma function
\\ \rowcolor{pastelpink}
$\exp\left(\cdot\right)$ & Exponential function
\\ \rowcolor{pastelpink}
$\det\left(\cdot\right)$ & Determinant operation
\\ \rowcolor{pastelpink}
$P\left(\cdot\right)$ & Probability of  an event.
\\ \rowcolor{pastelpink}
$N_{0}$ & Variance of AWGN noise
\\
\hline
\end{tabular}}
\label{TableListOfSymbols}
\end{table}

\section{System Model}
In this section, we present the working principle of the Alamouti space-time coded RIS-based RSM scheme depicted in Fig. \ref{Fig:SystemModel}. In our scheme, we propose the RIS as an access point (AP) that reflects the signals generated by a nearby radio frequency (RF) source thereby, transmission between the source and RIS is not affected by fading. Also, the destination, which consists of the $N_r$ active receive antennas, lies in the far-field of the RIS, so there is no line of side (LOS) between RF source and destination.

In the considered system model, bits generated from the source are split into two groups which have $\log_2(N_r)$ and $\log_2(M)$ bits, respectively. First $\log_2(N_r)$ bits determine the receive antenna index to adjust the RIS phases. The other $\log_2(M)$ bits are transmitted to the modulation block. After two symbols are selected in the modulation block, these symbols are transmitted to voltage-controlled oscillator  (VCO) by digital-to-analog converter (DAC). Signals carried to the IF by VCO reflected in the pre-defined receive antenna using RIS with $N$ reflector elements. Then, the received signals are brought to the baseband using a mixer and ow pass filter (LPF). The baseband signals are converted into digital signals, which are binary numbers. Lastly, we decide the received symbol and received antenna index according to digital output.

The RIS is equipped with $N$ passive reflecting elements. In this concept, we divide the RIS into two sub-surfaces, \textit{RIS-1} and \textit{RIS-2}, both of which have $N/2$ reflector elements. Each full transmission consists of two time slots in which we assume that the wireless fading channel does not change. The wireless fading channel between the receive antenna and $i^{th}$ reflector element in each sub-surface can be represented  as below;
\begin{equation}
    h_{l,i} = \beta_{l,i} \times e^{-j\Theta_{l,i}} \quad \quad l = 1,\cdots,N_r,
    \label{eq:channelDefination}
\end{equation}
where $\Theta_{l,i}$ is the channel phase induced by the $i^{th}$ reflector at the $l^{th}$ receive antenna, and $\beta_{l,i}$ is the channel fading coefficients followed Rayleigh distribution.

The controller determines the receiver antenna index according to $\log_2 N_r$ information bits in the first time slot. Using selected receiver antenna index, the channel phase of both sub-surfaces which are upper and lower are obtained as $\Theta_1 = \alpha_1$ and $\Theta_2 = \alpha_2$, respectively. Then, the RIS reflects the signal $s_1$ which is loaded into upper sub-surface \textit{RIS-1} and $s_2$ which is loaded into lower sub-surface \textit{RIS-2}. The base-band signal at $l^{th}$ receive antenna of the destination in the first time slot can be expressed as:
\begin{equation}
    y^{1}_{l} = s_1 \sum_{i = 1}^{N/2}h_{l,i}e^{j\Theta_{m,i}} + s_2 \sum_{i = \frac{N}{2}+1}^{N}h_{l,i}e^{j\Theta_{m,i}} + n^{1}_{l},
    \label{eq:RxSignalUpperRis}
\end{equation}
where $m \in \{1,2,\ldots,N_r\}$ and $n^1_{l}$ stand for the selected receive antenna and AWGN. The first and second terms come from upper sub-surface and lower sub-surface, respectively. We followed the same methods except transmitted symbols reflected from the RIS in the second time slot. We can represent the base-band signal at the $l^{th}$ receive antenna in the second time slot as below:
\begin{equation}
    y^{2}_{l} = -{s_2}^* \sum_{i = 1}^{N/2}h_{l,i}e^{j\Theta_{m,i}} + {s_1}^* \sum_{i = \frac{N}{2}+1}^{N}h_{l,i}e^{j\Theta_{m,i}} + n^2_{l}.
    \label{eq:RxSignalLowerRis}
\end{equation}
The spectral efficiency of RIS-RSM-ASBC system is $\eta = \left(2\times \log_2{M}+\log_2{N_r}\right)/2$ bits per channel use (bpcu) wheres traditional SM and RIS-RSM are $\log_2{M}+\log_2{N_r}$. It is clear that RIS-RSM-ASBC provides better spectral efficiency than classical SM schemes.

After combining  (\ref{eq:RxSignalUpperRis}) and (\ref{eq:RxSignalLowerRis}) into a matrix form, we get following:
\begin{equation}
\begin{aligned}
    \mathbf{y}_l &= \left[y^{1}_{l} \quad y^{2}_{l}\right]^T =
    \\
    &
    \underbrace{\left[\begin{array}{cc} s_1 & s_2 \\ -s_{2}^* & s_{1}^*
    \end{array}\right]}_{\mathbf{S}}
        \underbrace{\left[\begin{array}{c}     \sum_{i = 1}^{N/2}\beta_{l,i} e^{-j\Theta_{l,i}}e^{j\Theta_m,i} \\ \sum_{i = 1+N/2}^{N}\beta_{l,i} e^{-j\Theta_{l,i}}e^{j\Theta_m,i}
    \end{array}\right]}_{\mathbf{h}_l}
    +
   \underbrace{\left[\begin{array}{c} {n}_l^{1} \\ {n}_l^{2}    \end{array}\right]}_{\mathbf{n}_l}.
   \end{aligned}
    \label{eq:RxSignalCombined}
\end{equation}

At the receiver ML detector is utilized. We have applying ML into  (\ref{eq:RxSignalCombined}) to recover transmitted symbols and receiver antenna index:
\begin{equation}
    \Big(\hat{s}_1,\hat{s}_2,\hat{m}\Big) =  \arg \min\limits_{s_1,s_2,m}\sum\limits_{l=1}^{N_r}{\Big\vert\Big\vert \mathbf{y}_{l} - \mathbf{S}\mathbf{h}_{l}\Big\vert\Big\vert^2} ,
    \label{Estimated Signal}
\end{equation}
where $\hat{s}_1$, $\hat{s}_2$, $\hat{m}$ are estimation of ${s}_1$, ${s}_2$, ${m}$ respectively. 
\section{Performance Analysis}
In this section, theoretical average bit error rate (ABER) expression of the Alamouti coded RIS-based SM scheme are provided by evaluating pairwise error probability (PEP) and utilizing optimal ML detector. 

The conditional PEP expression of proposed system can be expressed as following:
\begin{eqnarray}\label{eq:equationPEP}
P\left(m,\mathbf{S}\rightarrow \hat{m},\hat{\mathbf{S}} \big\vert\mathbf{h}\right) \nonumber \\
&\!\!\!\!\!\!\!\!\!\!\!\!\!\!\!\!\!\!\!\!\!\!\!\!\!\!\!\!\!\!\!\!\!\!\!\!\!\!\!\!\!\!\!\!\!\!\!\!\!\!\!\!\!\!\!\!\!\!\!\!\!\!\!\!\!\!\!\!\!\!\!\!=&\!\!\!\!\!\!\!\!\!\!\!\!\!\!\!\!\!\!\!\!\!\!\!\!\!\!\!\!\!\!\!\!\!\!\!\!\!P\left(\sum_{l=1}^{N_r} \Big\vert \Big\vert \mathbf{y}_{l}-\mathbf{S}\mathbf{h}_{l}\Big\vert\Big\vert^2< \sum_{l=1}^{N_r}\Big\vert\Big\vert \mathbf{y}_{\hat{l}}-\mathbf{\hat {S}}\mathbf{h}_{\hat{l}}   \Big\vert\Big\vert^2 \right)\!.
\end{eqnarray}
Re-arranging (\ref{eq:equationPEP}) gives us the following:
\begin{eqnarray}
        P\left(m,\mathbf{S}\rightarrow \hat{m},\hat{\mathbf{S}} \big\vert\mathbf{h}\right)
     \nonumber
     \\
 &\!\!\!\!\!\!\!\!\!\!\!\!\!\!\!\!\!\!\!\!\!\!\!\!\!\!\!\!\!\!\!\!\!\!\!\!\!\!\!\!\!\!\!\!\!\!\!\!\!\!\!\!\!\!\!\!\!\!\!\!\!\!\!\!\!\!\!\!\!\!\!\!\!\!\!\!\!\!\!\!\!\!\!\!\!\!\!\!\!\!\!\!\!\!\!\!\!\!\!\!\!\!\!\!\!\!\!\!=&\!\!\!\!\!\!\!\!\!\!\!\!\!\!\!\!\!\!\!\!\!\!\!\!\!\!\!\!\!\!\!\!\!\!\!\!\!\!\!\!\!\!\!\!\!\!\!\!\!\!\!\!\!\!\!\!\! P\left(\sum_{l=1}^{N_r}\Big\vert\Big\vert \mathbf{n}_l\Big\vert\Big\vert^2 > \sum_{l=1}^{N_r} \Big\vert\Big\vert \mathbf{S}\mathbf{h}_l+ \mathbf{n}_{\hat{l}}-\mathbf{\hat{S}}\mathbf{h}_{\hat{l}}   \Big\vert\Big\vert^2\right)
     \nonumber\\
 &\!\!\!\!\!\!\!\!\!\!\!\!\!\!\!\!\!\!\!\!\!\!\!\!\!\!\!\!\!\!\!\!\!\!\!\!\!\!\!\!\!\!\!\!\!\!\!\!\!\!\!\!\!\!\!\!\!\!\!\!\!\!\!\!\!\!\!\!\!\!\!\!\!\!\!\!\!\!\!\!\!\!\!\!\!\!\!\!\!\!\!\!\!\!\!\!\!\!\!\!\!\!\!\!\!\!\!\!=&\!\!\!\!\!\!\!\!\!\!\!\!\!\!\!\!\!\!\!\!\!\!\!\!\!\!\!\!\!\!\!\!\!\!\!\!\!\!\!\!\!\!\!\!\!\!\!\!\!\!\!\!\!\!\!\!\!  P\left(\sum_{l=1}^{N_r}\Big\vert\Big\vert \mathbf{n}_l\Big\vert\Big\vert^2 - \left(\mathbf{S}\mathbf{h}_l-\mathbf{\hat{S}}\mathbf{h}_{\hat{l}} +\mathbf{n}_l \right)\right. \nonumber
\\
  &\!\!\!\!\!\!\!\!\!\!\!\!\!\!\!\!\!\!\!\!\!\!\!\!\!\!\!\!\!\!\!\!\!\!\!\!\!\!\!\!\!\!\!\!\!\!\!\!\!\!\!\!\!\!\!\!\!\!\!\!\!\!\!\!\!\!\!\!\!\!\!\!\!\!\!\!\!\!\!\!\!\!\!\!\!\!\!\!\!\!\!\!\!\!\!\!\!\!\!\!\!\!\!\!\!\!\!\!\times&\!\!\!\!\!\!\!\!\!\!\!\!\!\!\!\!\!\!\!\!\!\!\!\!\!\!\!\!\!\!\!\!\!\!\!\!\!\!\!\!\!\!\!\!\!\!\!\!\!\!\!\!\!\!\!\!\!\left.\left(\mathbf{S}\mathbf{h}_l-\mathbf{\hat{S}}\mathbf{h}_{\hat{l}} +\mathbf{n}_{\hat{l}} \right)^{H}>  0\right)
 \nonumber\\
 &\!\!\!\!\!\!\!\!\!\!\!\!\!\!\!\!\!\!\!\!\!\!\!\!\!\!\!\!\!\!\!\!\!\!\!\!\!\!\!\!\!\!\!\!\!\!\!\!\!\!\!\!\!\!\!\!\!\!\!\!\!\!\!\!\!\!\!\!\!\!\!\!\!\!\!\!\!\!\!\!\!\!\!\!\!\!\!\!\!\!\!\!\!\!\!\!\!\!\!\!\!\!\!\!\!\!\!\!=&\!\!\!\!\!\!\!\!\!\!\!\!\!\!\!\!\!\!\!\!\!\!\!\!\!\!\!\!\!\!\!\!\!\!\!\!\!\!\!\!\!\!\!\!\!\!\!\!\!\!\!\!\!\!\!\!\!P\Bigg(\underbrace{\sum_{l=1}^{N_r}-\Big\vert\Big\vert \mathbf{S}\mathbf{h}_l-\hat{\mathbf{S}}\mathbf{h}_{\hat{l}}\Big\vert\Big\vert^2 - 2\Re\left( \mathbf{n}_{\hat {l}}^{H} \left(\mathbf{S}\mathbf{h}_l-\mathbf{\hat{S}}\mathbf{h}_{\hat{l}}\right)\right)}_{\mathcal{K}}>  0\Bigg)
 \nonumber
 \\
\label{eq:qEquation}
\end{eqnarray}
where $\mathcal{K}$ is Gaussian R.V with mean $\mu_\mathcal{K} = -\sum_{l=1}^{N_r}\big\vert\big\vert \mathbf{S}\mathbf{h}_l-\hat{\mathbf{S}}\mathbf{h}_{\hat{l}}\big\vert\big\vert^2$ and variance $\sigma^2_{\mathcal{K}} = \sum_{l=1}^{N_r}2N_o\big\vert\big\vert \mathbf{S}\mathbf{h}_l-\hat{\mathbf{S}}\mathbf{h}_{\hat{l}}\big\vert\big\vert^2$. 

As known, $Q$ function can be defined as \cite{simon2008digital}:
\begin{equation}
P\left(X>x\right) = Q\left(x\right) = \frac{1}{2\pi}\int_{x}^{\infty}\exp\left({\frac{-u^2}{2}}\right)du,
\end{equation}
where $X \sim \mathcal{N}\left(0,1\right)$. 
So when we go back to our equation (\ref{eq:qEquation}), statistical values of $\mathcal{K}$ is re-arranged doing by normalization of the its standard deviation and shifting its mean to zero. Therefore the following expression is obtained:
\begin{equation}
\begin{aligned}
P\left(m,\mathbf{S}\rightarrow \hat{m},\hat{\mathbf{S}}\Big|\mathbf{h}\right) &=Q\left(\frac{-\mu_\mathcal{K}}{\sigma_\mathcal{K}}\right) 
\\
&=Q\left(\sqrt{\frac{\sum_{l=1}^{N_r}\Big\vert\Big\vert \mathbf{S}\mathbf{h}_{l}-\hat{\mathbf{S}}\mathbf{h}_{\hat{l}}\Big\vert\Big\vert^2}{2N_0}} \right).    
\end{aligned}
\end{equation}
Afterwards, the unconditional PEP expression can be obtained taking by the expectation of (\ref{eq:equationPEP}):
\begin{multline}
\overline{P}_e\left(m,\mathbf{S}\rightarrow \hat{m},\hat{\mathbf{S}}\right) =  \int_0^{\infty} Q\left(\sqrt{\frac{\Gamma}{2N_0}}\right)f_{\Gamma}(\Gamma)
\\
= \frac{1}{\pi}\int_0^{\infty}\int_0^{\pi/2}\exp{\left(-\frac{\Gamma}{4\sin^2{\Theta}N_0}\right) f_\Gamma\left(\Gamma\right)d\Theta d\Gamma }
\\
\overline{P}_e\left(m,\mathbf{S}\rightarrow \hat{m},\hat{\mathbf{S}}\right) =\frac{1}{\pi}\int_0^{\pi/2}M_\Gamma\left(-\frac{1}{4\sin^2{\Theta}N_0}\right),
\label{eq:MGFequation}
\end{multline}
where  $\Gamma =\sum_{l=1}^{N_r}\Big\vert\Big\vert \mathbf{S}\mathbf{h}_l-\hat{\mathbf{S}}\mathbf{h}_{\hat{l}}\Big\vert\Big\vert^2$. 

We need the MGF of $\Gamma$ $(M_\Gamma(s))$ to perform this integration. So we will divide the R.V $\Gamma$ into two parts. From this point of view, we can re-write the $\Gamma$ more precisely as follows:
\begin{equation}
 \Gamma = \Gamma_{Ts_1} + \Gamma_{Ts_2},
    \label{eq:GammaExpansion}
\end{equation}
where
\begin{equation}
\begin{aligned}
    \Gamma_{Ts_1} &=  \sum_{l=1}^{Nr} \left\vert \sum_{i=1}^{N/2}\beta_{l,i}e^{-j\Theta_{l,i}}\left(e^{j\Theta_{m,i}}s_1  -e^{j\Theta_{\hat{m},i}}\hat{s}_1  \right) \right.
    \\
    &+\left.
    \sum_{i=1+N/2}^{N}\beta_{l,i} e^{-j\Theta_{l,i}}\left(e^{j\Theta_{m,i}}s_2 -e^{j\Theta_{\hat{m},i}}\hat{s}_2  \right)
    \right\vert^2,
    \\
    \Gamma_{Ts_2} &=  \sum_{l=1}^{Nr} \left\vert- \sum_{i=1}^{N/2}\beta_{l,i}e^{-j\Theta_{l,i}}\left(e^{j\Theta_{m,i}}s_2^{*} -e^{j\Theta_{\hat{m},i}}\hat{s}_2^{*}  \right) \right.
    \\
    &+\left.
    \sum_{i=1+N/2}^{N}\beta_{l,i} e^{-j\Theta_{l,i}}\left(e^{j\Theta_{m,i}}s_1^{*} -e^{j\Theta_{\hat{m},i}}\hat{s}_1^{*}  \right)
    \right\vert^2.
\end{aligned}
\end{equation}

The point to be noted in this expression is the fact that the $\Gamma_{Ts_1}$ and $\Gamma_{Ts_2}$ are independent of each other due to the fact that signals are experienced with different time zones. Plus, $\Gamma_{Ts_1} \delequal \Gamma_1$ and $\Gamma_{Ts_2}\delequal\Gamma_2$  will be expressed as for the simplicity. Therefore, separate MGF derivations will be made for two time slots.

\subsection{The MGFs Derivation for First Time Slot}
This MGF can be derived by considering the general quadratic form of correlated Gaussian RVs and depends on erroneous or correct detection of the receive antenna index $m$.

\subsubsection{Under the wrong antenna decision, $m\neq\hat{m}$}

\definecolor{aureolin}{rgb}{0.99, 0.99, 0}
\definecolor{arylideyellow}{rgb}{1, 0.99, 0.7}

\begin{table}
  \caption{The expressions of $\Gamma_{11},\Gamma_{12},\Gamma_{13}, \Gamma_{21}$, $\Gamma_{22}$ and $\Gamma_{23}$ (First row: Time Slot-1 and Second row: Time Slot-2).}
\centering
\begin{tabular}{l c}
\hline\hline 
\rowcolor{arylideyellow}
  $ \begin{aligned}
\Gamma_{11} &=  \left\vert\sum_{i=1}^{N/2}\beta_{m,i}\left(s_1-e^{-j\Psi_i} \hat{s}_1\right) + \!\!\!\!\!\!\!\sum_{i=1+N/2}^{N}\!\!\!\!\!\beta_{m,i}\left(s_2-e^{-j\Psi_i}\hat{s}_2\right)\right\vert^2
\\
\Gamma_{12} &=  \left\vert\sum_{i=1}^{N/2}\beta_{\hat{m},i}\left(s_{1}e^{-j\Psi_i} -\hat{s}_1\right) + \!\!\!\!\!\!\!\sum_{i=1+N/2}^{N}\!\!\!\!\!\beta_{\hat{m},i}\left(s_{2}e^{-j\Psi_i} -\hat{s}_2\right)\right\vert^2
\\
\Gamma_{13} &= \sum_{l=1(l\neq m,l\neq \hat{m})}^{N_r}\Bigg\vert\sum_{i=1}^{N/2}\beta_{l,i}\left(s_{1}e^{-j\Theta_{l,i}} -\hat{s}_{1}e^{-j\Theta_{\hat{m},i}}\right) \\
&+ \sum_{i=1+N/2}^{N}\beta_{l,i}\left(s_{2}e^{-j\Theta_{l,i}} -\hat{s}_{2}e^{-j\Theta_{\hat{m},i}}\right)\Bigg\vert^2 
\end{aligned}$
    \\
    \hline \rowcolor{arylideyellow}
    $
    \begin{aligned}
    \\
\Gamma_{21} &=  \left\vert\sum_{i=1}^{N/2}\beta_{m,i}\left(-s_2^{*}+e^{-j\Psi_i} \hat{s}_2^{*}\right) + \!\!\!\!\!\!\!\sum_{i=1+N/2}^{N}\!\!\!\!\!\beta_{m,i}\left(s_1^{*}-e^{-j\Psi_i}\hat{s}_1^{*}\right)\right\vert^2
\\
\Gamma_{22} &=  \left\vert\sum_{i=1}^{N/2}\beta_{\hat{m},i}\left(\hat{s}^{*}_2 - s_{2}^{*}e^{-j\Psi_i}\right) + \!\!\!\!\!\!\!\sum_{i=1+N/2}^{N}\!\!\!\!\!\beta_{\hat{m},i}\left(s_{1}^{*}e^{-j\Psi_i} -\hat{s}_1^{*}\right)\right\vert^2
\\
\Gamma_{23} &= \sum_{l=1(l\neq m,l\neq \hat{m})}^{N_r}\Bigg \vert\sum_{i=1}^{N/2}\beta_{l,i}\left(s_{1}^{*}e^{-j\Theta_{m,i}} -\hat{s}_{1}^{*}e^{-j\Theta_{\hat{m},i}}\right) \\
&+ \sum_{i=1+N/2}^{N}\beta_{l,i}\left(s_{2}^{*}e^{-j\Theta_{m,i}} -\hat{s}_{2}^{*}e^{-j\Theta_{\hat{m},i}}\right)\Bigg\vert^2 
\end{aligned}
$
\\
\hline
\hline
\end{tabular}
\label{TableGammaExpressions}
\end{table}

We can re-write $\Gamma_1$ as $\Gamma_1 =\Gamma_{11} + \Gamma_{12} + \Gamma_{13}$ and these R.Vs are defined in Table \ref{TableGammaExpressions}. In this manner, $\Gamma_{11}$, $\Gamma_{12}$ and $\Gamma_{13}$ stand for $l=m$, $l=\hat{m}$ and $l\neq\hat{m},l\neq m$. Also, $ \Psi_i = \Theta_{m,i}-\Theta_{\hat{m},i}$ is with having triangular distribution \cite{petrov2012sums}:
\begin{equation}
 f_{\Psi}(x)=\left\{ \begin{array}{rcl}
{2\pi}\left(1+\frac{x}{2\pi}\right);& -2\pi<x<0
\\{2\pi}\left(1-\frac{x}{2\pi}\right);& 0<x<2\pi
\end{array}\right.   
\end{equation}

Here, we will firstly derive the MGF expression of $\Gamma_{11} + \Gamma_{12}$. After that, MGF of $\Gamma_{13}$ will be presented. It is known that the MGF of the sum of two independent R.Vs is equal to the product of their MGFs. Hence, we will obtain MGF of $\Gamma_1$ as $M_{\Gamma_{1}} = M_{\left(\Gamma_{11}+\Gamma_{12}\right)}\times M_{\Gamma_{13}}$.

\definecolor{maroon}{cmyk}{0,0.87,0.68,0.32}

\begin{table}
  \caption{The derivation of entities of $\mathbf{m}$ and $\mathbf{\Tilde{\mathbf{m}}}$, (First row: Time Slot-1 and  Second row: Time Slot-2)}
\centering

\begin{tabular}{l c}
\hline\hline  \rowcolor{maroon!10}
  $ \begin{aligned}
        \mu_{\left(\Delta_{11}\right)_\Re} &= \frac{N\sqrt{\pi}}{4}\Big((s_1)_\Re + (s_2)_\Re\Big), 
    \\
       \mu_{\left(\Delta_{11}\right)_\Im} &= \frac{N\sqrt{\pi}}{4}\Big((s_1)_\Im + (s_2)_\Im\Big),
    \\
    \mu_{\left(\Delta_{12}\right)_\Re} &= -\frac{N\sqrt{\pi}}{4}\Big(\left(\hat{s}_1\right)_\Re + (\hat{s}_2)_\Re\Big)
    
        \\
    \mu_{\left(\Delta_{12}\right)_\Im} &= -\frac{N\sqrt{\pi}}{4}\Big((\hat{s}_1)_\Re + (\hat{s}_2)_\Re\Big)
    \\
    \\
    \hline
    \\   
     \mu_{\left(\Delta_{21}\right)_\Re} &= \frac{N\sqrt{\pi}}{4}\Big((s^*_1)_\Re - (s^*_2)_\Re\Big)
    \\   
       \mu_{\left(\Delta_{21}\right)_\Im} &= \frac{N\sqrt{\pi}}{4}\Big((s^*_1)_\Im - (s^*_2)_\Im\Big),
    \\
    \mu_{\left(\Delta_{22}\right)_\Re} &= -\frac{N\sqrt{\pi}}{4}\Big(\left(\hat{s}^*_1\right)_\Re - (\hat{s}^*_2)_\Re\Big)
    \\
    \mu_{\left(\Delta_{22}\right)_\Im} &= -\frac{N\sqrt{\pi}}{4}\Big((\hat{s}^*_1)_\Re - (\hat{s}^*_2)_\Re\Big)    
    \end{aligned}$
    \\
    \hline
    \hline
\end{tabular}
\label{TableMeanM1}
\end{table}

Keeping in mind, $\Gamma_{11} = \vert\Delta_{11}\vert^2 = \left({\Delta_{11}}\right)^2_{\Re} + \left({\Delta_{11}}\right)^2_{\Im}$ , $\Gamma_{12} = \vert\Delta_{12}\vert^2 = \left({\Delta_{12}}\right)^2_{\Re} + \left({\Delta_{12}}\right)^2_{\Im}$  and $\Gamma_{11}+\Gamma_{12} = \mathbf{x}^T\mathbf{A}\mathbf{x}$ where 
\begin{equation}
\begin{aligned}
    \mathbf{x} &= \Big[\begin{array}{cccc}\left(\Gamma_{11}\right)_\Re & \left(\Gamma_{11}\right)_\Im & \left(\Gamma_{12}\right)_\Re & \left(\Gamma_{12}\right)_\Im\end{array}\Big]^{T},
    \\
    \mathbf{A} &=  \begin{bmatrix}
       1 & 0 & 0 & 0 
       \\
       0 & 1 & 0 & 0 
       \\
       0 & 0 & 1 & 0 
       \\ 
      0 & 0 & 0 & 1
      \end{bmatrix}
      \\
      \mathbf{x}^T\mathbf{A}\mathbf{x} &= \left({\Delta_{11}}\right)^2_{\Re} + \left({\Delta_{11}}\right)^2_{\Im} + \left({\Delta_{12}}\right)^2_{\Re} + \left({\Delta_{12}}\right)^2_{\Im}
\end{aligned}
\end{equation}
Let $\mathcal{G}= \Gamma_{11}+\Gamma_{12}$, then MGF of $\mathcal{G}$ can be obtained as follows \cite[Eq. (3.2a.1)]{mathai1992quadratic}:
\begin{equation}
\begin{aligned}
M_{\mathcal{G}}(s)&=\left(\sqrt{\textrm{det}\left(\mathbf{I}-2s\mathbf{AC}\right)}\right)^{-1}
\\
&\times\exp\left(-\frac{1}{2}\mathbf{m}^T\left[\mathbf{I}-\left(\mathbf{I}-2s\mathbf{AC}\right)^{-1}\right]\mathbf{C}^{-1}\right).
\end{aligned}
\end{equation}
where $\mathbf{m}$ and $\mathbf{C}$ are the mean vector and the covariance matrix of
$\mathbf{x}$, respectively.
We can define mean vector of $\mathbf{x}$ by $\mathbf{m}=\left[\begin{array}{cccc}\mu_{\left(\Delta_{11}\right)_\Re} & \mu_{\left(\Delta_{11}\right)_\Im} &\mu_{\left(\Delta_{12}\right)_\Re} & \mu_{\left(\Delta_{12}\right)_\Im}\end{array}\right]^T $. The entities of $\mathbf{m}$ can be represented in Table \ref{TableMeanM1}.

\definecolor{LightCyan}{rgb}{0.88,1,1}
\begin{table*}[t]
  \caption{Statistical Properties of the covariance entities of $\mathbf{C}$ and $\Tilde{\mathbf{C}}$ (First row: Time Slot-1 and Second row: Time Slot-2) }
\centering

\begin{tabular}{l c}
\hline\hline 
 \rowcolor{LightCyan}
 $
    \begin{aligned}
\sigma_{1}^2 &= \frac{N}{8}\left(4-\pi\right)\Big((s_1)_{\Re}^2 +(s_2)_{\Re}^2\Big) + \frac{N}{4}\Big(\vert\hat{s}_1\vert^2 + \vert\hat{s}_2\vert^2 \Big),\; \sigma_{1,3} = \frac{-N\pi}{16}\Big((s_1)_{\Re}(\hat{s}_1)_{\Re} +(s_1)_{\Im}(\hat{s}_1)_{\Im} +(s_2)_{\Re}(\hat{s}_2)_{\Re} + (s_2)_{\Im}(\hat{s}_2)_{\Im}\Big)
\\
\sigma_{2}^2 &= \frac{N}{8}\left(4-\pi\right)\Big((s_1)_{\Im}^2 +(s_2)_{\Im}^2\Big) + \frac{N}{4}\Big(\vert\hat{s}_1\vert^2 + \vert\hat{s}_2\vert^2 \Big),\; \sigma_{1,4} = \frac{-N\pi}{16}\Big((s_1)_{\Im}(\hat{s}_1)_{\Re} +(s_1)_{\Re}(\hat{s}_1)_{\Im} +(s_2)_{\Im}(\hat{s}_2)_{\Re} + (s_2)_{\Re}(\hat{s}_2)_{\Im}\Big)
\\
\sigma_{3}^2 &= \frac{N}{8}\left(4-\pi\right)\Big((\hat{s}_1)_{\Re}^2 +(\hat{s}_2)_{\Re}^2\Big) + \frac{N}{4}\Big(\vert {s}_1\vert^2 + \vert {s}_2\vert^2 \Big), \; \sigma_{2,4} = \frac{N\pi}{16}\Big((s_1)_{\Re}(\hat{s}_1)_{\Re} +(s_1)_{\Im}(\hat{s}_1)_{\Im} +(s_2)_{\Re}(\hat{s}_2)_{\Re} + (s_2)_{\Im}(\hat{s}_2)_{\Im}\Big)
\\
\sigma_{4}^2 &= \frac{N}{8}\left(4-\pi\right)\Big((s_1)_{\Im}^2 +(s_2)_{\Im}^2\Big) + \frac{N}{4}\Big(\vert{s}_1\vert^2 + \vert{s}_2\vert^2 \Big),\;  \sigma_{2,3} = \frac{-N\pi}{16}\Big((\hat{s}_1)_{\Re}(s_1)_{\Im} +(s_1)_{\Re}(\hat{s}_1)_{\Im} +(s_2)_{\Im}(\hat{s}_2)_{\Re} + (s_2)_{\Re}(\hat{s}_2)_{\Im}\Big)
\\
\sigma_{1,2} &= \frac{N(4-\pi)}{8}\Big((s_1)_{\Re}(s_1)_{\Im} + (s_2)_{\Re}(s_2)_{\Im}\Big), \;\;\;\;\;\;\;\;\;\;\;\;\;\;\;\; \sigma_{3,4} = \frac{N(4-\pi)}{8}\Big((\hat{s}_1)_{\Re}(\hat{s}_1)_{\Im} + (\hat{s}_2)_{\Re}(\hat{s}_2)_{\Im}\Big)
\\
\end{aligned}
$
\\
\hline
\rowcolor{LightCyan}
$ 
 \begin{aligned}
\sigma_{1}^2 &= \frac{N}{8}\left(4-\pi\right)\Big((s^*_1)_{\Re}^2 +(s^*_2)_{\Re}^2\Big) + \frac{N}{4}\Big(\vert\hat{s}^*_1\vert^2 + \vert\hat{s}^*_2\vert^2 \Big),\; \sigma_{1,3} = \frac{N\pi}{16}\Big((s^*_1)_{\Im}(\hat{s}^*_1)_{\Im}-(s^*_1)_{\Re}(\hat{s}^*_1)_{\Re} -(s^*_2)_{\Re}(\hat{s}^*_2)_{\Re} + (s^*_2)_{\Im}(\hat{s}^*_2)_{\Im}\Big)
\\
\sigma_{2}^2 &= \frac{N}{8}\left(4-\pi\right)\Big((s^*_1)_{\Im}^2 +(s^*_2)_{\Im}^2\Big) + \frac{N}{4}\Big(\vert\hat{s}^*_1\vert^2 + \vert\hat{s}^*_2\vert^2 \Big),\; \sigma_{1,4} = \frac{-N\pi}{16}\Big((s^*_1)_{\Im}(\hat{s}^*_1)_{\Re} +(s^*_1)_{\Re}(\hat{s}^*_1)_{\Im} +(s^*_2)_{\Im}(\hat{s}^*_2)_{\Re} + (s^*_2)_{\Re}(\hat{s}^*_2)_{\Im}\Big)
\\
\sigma_{3}^2 &= \frac{N}{8}\left(4-\pi\right)\Big((\hat{s}^*_1)_{\Re}^2 +(\hat{s}^*_2)_{\Re}^2\Big) + \frac{N}{4}\Big(\vert {s}^*_1\vert^2 + \vert {s}^*_2\vert^2 \Big),\; \sigma_{2,4} = \frac{N\pi}{16}\Big((s^*_1)_{\Re}(\hat{s}^*_1)_{\Re} -(s^*_1)_{\Im}(\hat{s}^*_1)_{\Im} +(s^*_2)_{\Re}(\hat{s}^*_2)_{\Re} - (s^*_2)_{\Im}(\hat{s}^*_2)_{\Im}\Big)
\\
\sigma_{4}^2 &= \frac{N}{8}\left(4-\pi\right)\Big((s^*_1)_{\Im}^2 +(s^*_2)_{\Im}^2\Big) + \frac{N}{4}\Big(\vert{s}^*_1\vert^2 + \vert{s}^*_2\vert^2 \Big),\; \sigma_{2,3} = \frac{-N\pi}{16}\Big((\hat{s}^*_1)_{\Re}({s}^*_1)_{\Im} +(s^*_1)_{\Re}(\hat{s}^*_1)_{\Im} +(s^*_2)_{\Im}(\hat{s}^*_2)_{\Re} + (s^*_2)_{\Re}(\hat{s}^*_2)_{\Im}\Big)
\\
\sigma_{1,2} &= \frac{N(4-\pi)}{8}\Big((s^*_1)_{\Re}(s^*_1)_{\Im} + (s^*_2)_{\Re}(s^*_2)_{\Im}\Big),\;\;\;\;\;\;\;\;\;\;\;\;\;\;\;\; \sigma_{3,4} = \frac{N(4-\pi)}{8}\Big((\hat{s}^*_1)_{\Re}(\hat{s}^*_1)_{\Im} + (\hat{s}^*_2)_{\Re}(\hat{s}^*_2)_{\Im}\Big)
\\
\end{aligned} $
\\
\hline
\hline
\end{tabular}
\label{CoavarainaceEntitesX1}
\end{table*}

Likewise, we can give the covariance matrix of $\mathbf{x}$ as below:

\begin{equation}
\begin{aligned}
    \mathbf{C} &=  \begin{bmatrix}
       \sigma_{1}^2 & \sigma_{1,2} & \sigma_{1,3} & \sigma_{1,4} 
       \\
       \sigma_{1,2} & \sigma_{2}^2 & \sigma_{2,3} & \sigma_{2,4}  
       \\
       \sigma_{1,3}& \sigma_{2,3} & \sigma_{3}^2 & \sigma_{3,4} 
       \\ 
        \sigma_{1,4}  & \sigma_{2,4}   & \sigma_{3,4} & \sigma_{4}^2
      \end{bmatrix}.
\end{aligned}    
\end{equation}
Hence, the entities of covariance matrix are obtained as in Table \ref{CoavarainaceEntitesX1} (at the top of the next page). On the other hand, the MGF of the $\Gamma_{13}$ is found. $\Gamma_{13}$ in Table \ref{TableGammaExpressions} can be rewritten as follows:
\begin{equation}
\Gamma_{13} = \sum_{l=1(l\neq m,l\neq \hat{m})}^{N_r}\Bigg\vert\sum_{i=1}^{N/2}\beta_{l,i}s_{i,1} + \sum_{i=1+N/2}^{N}\beta_{l,i}s_{i,2}\Bigg\vert^2,     
\end{equation}
where $s_{i,1} =\left(s_{1}e^{-j\Theta_{l,i}} -\hat{s}_{1}e^{-j\Theta_{\hat{m},i}}\right) $ and $s_{i,2}= \left(s_{2}e^{-j\Theta_{l,i}} -\hat{s}_{2}e^{-j\Theta_{\hat{m},i}}\right)$. Thanks to the variables $s_{i,1}$ and $s_{i,2}$ are independent and due to the CLT for the large $N$ ($N\gg1$), the $\Gamma_{13}$ expression have a central zero mean Chi-Squared distribution with order $(N_r-2)$. Keep in mind that $X=\sum_{i=1}^{n}Z_i$ where $Z_i\sim \mathcal{N}\left(0,\sigma_{Z}^2\right)$, the MGF of $X$ with a central zero-mean Chi-Square distribution can be written as \cite{simon2002probability} :
\begin{equation}
\begin{aligned}
M_{X}(s)=\left(\frac{1}{1-2s\sigma^2_{X}}\right)^{n/2},
\label{Chi-squaredMGF}
\end{aligned}
\end{equation}
where $n$ stands for degree of freedom. Therefore, we can write the MGF of $\Gamma_{13}$ by appropriately changing the parameters in the equation (\ref{Chi-squaredMGF}) :
\begin{equation}
  M_{\Gamma_{13}}(s)=  \left(\frac{1}{1- \frac{Ns}{2}\left(\vert{s_1}\vert^2 +\vert{\hat{s}_1}\vert^2+\vert{s_2}\vert^2 +\vert{\hat{s}_2}\vert^2\right)}\right)^{N_r-2}.
  \label{Gamma13 Expression}
\end{equation}
Eventually,  the MGF of $\Gamma_{1}$ in (\ref{eq:GammaExpansion}) is yielded by the product of MGFs of $\Gamma_{11}+\Gamma_{12}$ and $\Gamma_{13}$. 

\subsubsection{Under the correct antenna decision, $\hat{m} = m$}
In this case, we will calculate the PEP for correct antenna detection. We can write $\Gamma_1$ for $\hat{m} = m$ as follows:
\begin{equation}
\begin{aligned}
    \Gamma^{\left(\hat{m} = m\right)}_{1} &=  \sum_{l=1}^{Nr} \left\vert \sum_{i=1}^{N/2}\beta_{l,i}e^{j\left(\Theta_{l,i}-\Theta_{m,i}\right)}\left(s_1  -\hat{s}_1  \right) \right.
    \\
    &+\left.
    \sum_{i=1+N/2}^{N}\beta_{l,i}e^{j\left(\Theta_{l,i}-\Theta_{m,i}\right)}\left(s_2 -\hat{s}_2 \right)
    \right\vert^2
    \\
    \Gamma_{1}^{\left(\hat{m} = m\right)} &=\sum_{l=1}^{Nr} \left\vert H_{l,i}\left(s_1  -\hat{s}_1  \right)+
    H_{l,i}\left(s_2 -\hat{s}_2 \right)
    \right\vert^2
    \\
    &\leq \sum_{l=1}^{Nr} \left\vert H_{l,i}\left(s_1  -\hat{s}_1  \right)\right\vert^2
    +
     \sum_{l=1}^{Nr} \left\vert H_{l,i}\left(s_2 -\hat{s}_2 \right)
    \right\vert^2,   
\end{aligned}
\label{GammaTs1CorrectAntenna}
\end{equation}
where $H_{l,i} =\sum_{i=1}^{N/2}\beta_{l,i}e^{j\left(\Theta_{l,i}-\Theta_{m,i}\right)} $. It is worth mentioning that we utilize the upper bound approach in (\ref{GammaTs1CorrectAntenna}). Let's suppose that $A=\sum_{l=1}^{Nr} \left\vert H_{l,i}\left(s_1  -\hat{s}_1  \right)\right\vert^2 $ and $B = \sum_{l=1}^{Nr}  \left\vert H_{l,i}\left(s_2 -\hat{s}_2 \right)
\right\vert^2$, the MGF of $M_{\Gamma_{1}}(s) = M_{A}(s)\times M_{B}(s)$. We can re-write clearly $A_{1}$ and $B_{1}$ as follows:
\begin{equation}
\begin{aligned}
    A_{1}&=  \left(s_1  -\hat{s}_1  \right)\left(\left\vert H_{m,i}\right\vert^2 + \sum_{l\neq m}^{Nr}\left\vert H_{l,i}\right\vert^2\right),
    \\
    B_{1}&=\left(s_2  -\hat{s}_2  \right)\left(\left\vert H_{m,i}\right\vert^2 + \sum_{l\neq m}^{Nr}\left\vert H_{l,i}\right\vert^2\right),
\end{aligned}
\end{equation}
where $ H_{m,i} = \sum_{i=1}^{N/2}\beta_{m,i}$ and distributed as $ H_{m,i} \sim \mathcal{N}\left(\frac{\sqrt{\pi}N}{4},\frac{\left(4-\pi\right)N}{8}\right)$  for large $N$ (due to the Central Limit Theorem, (CLT)). Similarly,  $ H_{m,i} \sim \mathcal{N}\left(0,\frac{N}{2}\right)$. Therefore, the MGFs of $A$ and $B$ can be obtained product of two $\chi^2$ variables respectively as follows \cite{simon2002probability}:
\begin{equation}
\begin{aligned}
M_{A_1}(s)&= \left(\frac{1}{1-\frac{sN(4-\pi)\vert{s_1-\hat{s}_1}\vert^2}{4}}\right)^{1/2}
\\
\times&\exp{\left(\frac{\frac{sN^2\pi\vert{s_1-\hat{s}_1}\vert^2}{16} }{1 -\frac{sN(4-\pi)\vert{s_1-\hat{s}_1}\vert^2}{4}} \right)}\left(\frac{1}{1-\frac{sN\vert{s_1-\hat{s}_1}\vert^2}{2}}\right)^{N_r-1}
\\
M_{B_1}(s)&= \left(\frac{1}{1-\frac{sN(4-\pi)\vert{s_2-\hat{s}_2}\vert^2}{4}}\right)^{1/2}
\\
\times&\exp{\left(\frac{\frac{sN^2\pi\vert{s_2-\hat{s}_2}\vert^2}{16} }{1 -\frac{sN(4-\pi)\vert{s_2-\hat{s}_2}\vert^2}{4}} \right)}\left(\frac{1}{1-\frac{sN\vert{s_2-\hat{s}_2}\vert^2}{2}}\right)^{N_r-1}.
\end{aligned}
\end{equation}
Finally, the MGF of $\Gamma_1^{\left(\hat{m} = m\right)}$ is yielded from the product of MGFs of $A_1$ and $B_1$ for correct antenna detection. 

\subsection{The MGFs Derivation for Second Time Slot}
In light of this information, the analyses in Section 3.A (for
the case of both $m \neq\hat{m}$ and $m = \hat{m}$ ) is also valid for Second Time Slot. Therefore same procedures are applied in this section.
\subsubsection{Under the wrong antenna decision, $\hat{m} \neq m$}
The MGF of $\Gamma_{21}$, $\Gamma_{22}$ and $\Gamma_{23}$ derived from (12) with suitable modifications in (14-15), (19).
Covariance ($\Tilde{\mathbf{C}}$) and mean ($\Tilde{\mathbf{x}}$) entites  are given for MGF of $\left(\Gamma_{21}+\Gamma_{22}\right)$ in Table IV and Table V respectively. Also, the MGF of $\Gamma_{23}$ can be given by doing proper modifications in (\ref{Gamma13 Expression}) as following :
\begin{equation}
  M_{\Gamma_{23}}(s)=  \left(\frac{1}{1- \frac{Ns}{2}\left(\vert{s^*_1}\vert^2 +\vert{\hat{s}^{*}_1}\vert^2+\vert{s^{*}_2}\vert^2 +\vert{\hat{s}^{*}_2}\vert^2\right)}\right)^{N_r-2},
  \label{Gamma23 Expression}
\end{equation}

\subsubsection{Under the correct antenna decision, $\hat{m} = m$}
The analyses in Section 3.A.2 is also applicable for this case. Firstly, we can write MGF of $A_{2}$ and $B_{2}$ as below:

\begin{equation}
\begin{aligned}
M_{A_2}(s)&= \left(\frac{1}{1-\frac{sN(4-\pi)\vert{\hat{s}^*_2-s^*_2}\vert^2}{4}}\right)^{1/2}
\\
\times&\exp{\left(\frac{\frac{sN^2\pi\vert{\hat{s}^*_2-s^*_2}\vert^2}{16} }{1 -\frac{sN(4-\pi)\vert{\hat{s}^*_2-s^*_2}\vert^2}{4}} \right)}\left(\frac{1}{1-\frac{sN\vert{\hat{s}^{*}_2-s^{*}_2}\vert^2}{2}}\right)^{N_r-1}
\\
M_{B_2}(s)&= \left(\frac{1}{1-\frac{sN(4-\pi)\vert{s^*_1-\hat{s}^{*}_1}\vert^2}{4}}\right)^{1/2}
\\
\times&\exp{\left(\frac{\frac{sN^2\pi\vert{s^{*}_1-\hat{s}^{*}_1}\vert^2}{16} }{1 -\frac{sN(4-\pi)\vert{s^{*}_1-\hat{s}^{*}_1}\vert^2}{4}} \right)}\left(\frac{1}{1-\frac{sN\vert{s^{*}_1-\hat{s}^{*}_1}\vert^2}{2}}\right)^{N_r-1}
\end{aligned}
\label{}
\end{equation}
In this manner, MGF of $\Gamma^{\left(\hat{m} = m\right)}_{23}$ can be obtained as $\Gamma^{\left(\hat{m} = m\right)}_{23} = M_{A_2}(s)\times M_{B_2}(s)$.  

Thus, separate MGF derivations have been made for two time slots containing wrong and correct antenna detection cases. The PEP values for each case are obtained by substituting and integrating the corresponding MGF expression in (\ref{eq:MGFequation}).

Finally, the bit error probability (BEP) upper bound of RIS-RSM-ASBC can be expressed as follows:
\begin{equation}
P_{b}\leq\frac{1}{2^{2\eta}}\sum_{\mathbf{S},\mathbf{\hat{S}}}\sum_{m,\hat{m}}{\frac{P\left(m,\mathbf{S}\rightarrow \hat{m},\mathbf{\hat{S}}\right)e\left(m,\mathbf{S},\hat{m},\mathbf{\hat{S}}\right)}{2\eta}},
\end{equation}
where, $P\left(m,\mathbf{S}\rightarrow \hat{m},\mathbf{\hat{S}}\right)$ and $e\left(m,\mathbf{S},\hat{m},\mathbf{\hat{S}}\right)$ stand for the PEP and the number of bit errors associated with the corresponding PEP events, respectively.

\section{Numerical Results and Complexity Analysis}

In this section, first the numerical results of the proposed system and then the system complexity analysis will be presented.
\subsection{Numerical Results}
Here, we provide the Monte-Carlo simulations for the proposed Alamouti Coded RIS-Based Received SM and compare them with the theoretical results obtained. ML detector is utilized to detect the transmitted symbols and indices on the receiving side. We assume that all fading channels are uncorrelated Rayleigh distribution. The SNR parameter used in the simulations  is expressed as: $\mathrm{SNR (dB)}=10\log_{10}(E_s/N_0)$, where $E_s$ is the average symbol energy.

\begin{figure}
     \centering
\includegraphics[width=0.5\textwidth]{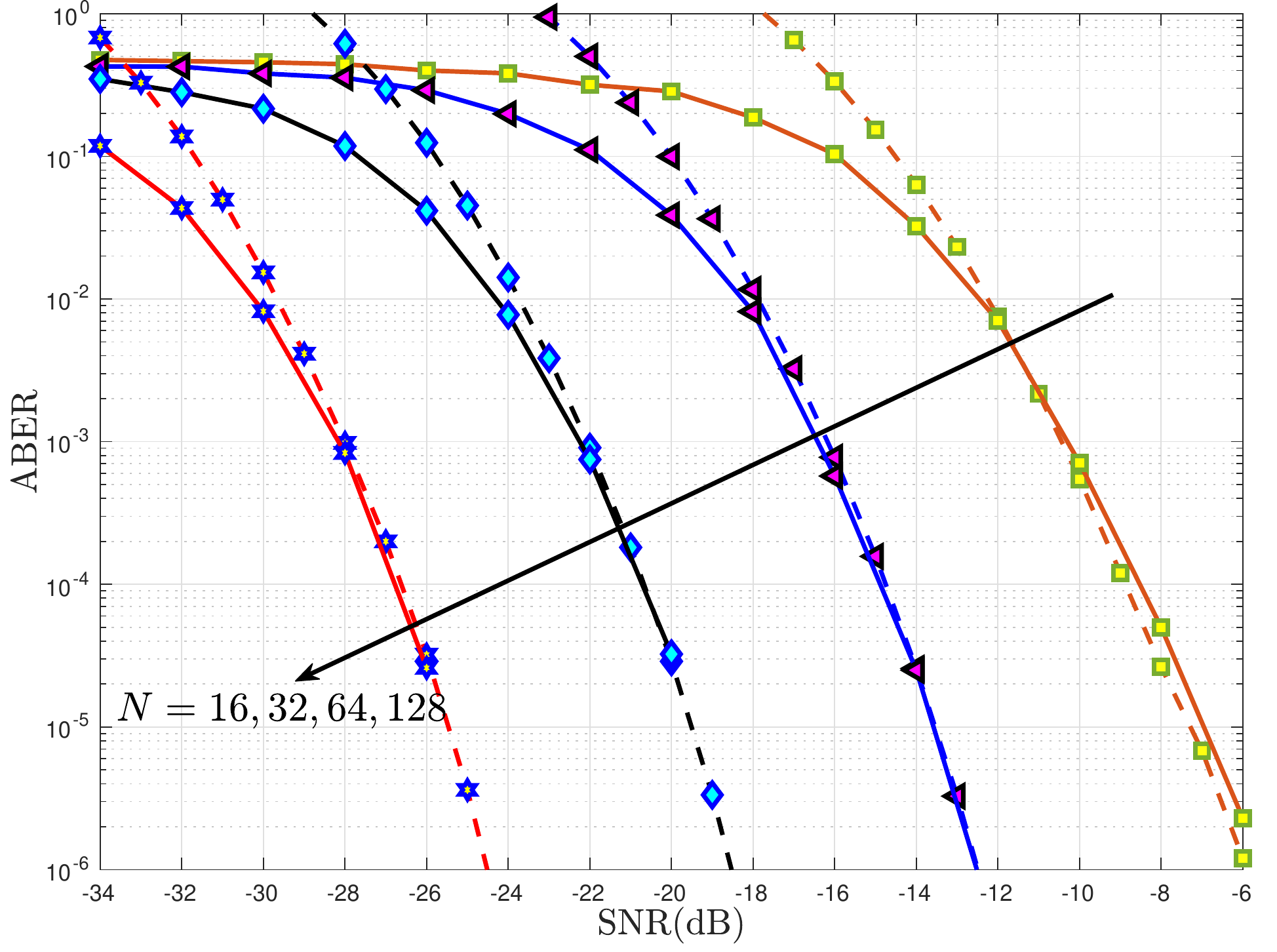}
        \caption{BER performance of the ASBC coded RIS based Received SM with $(M=4,N_r=4)$ for varying $N$.  }
        \label{fig:ProposedSchemeBER}
\end{figure}

\begin{figure}
     \centering
\includegraphics[width=0.5\textwidth]{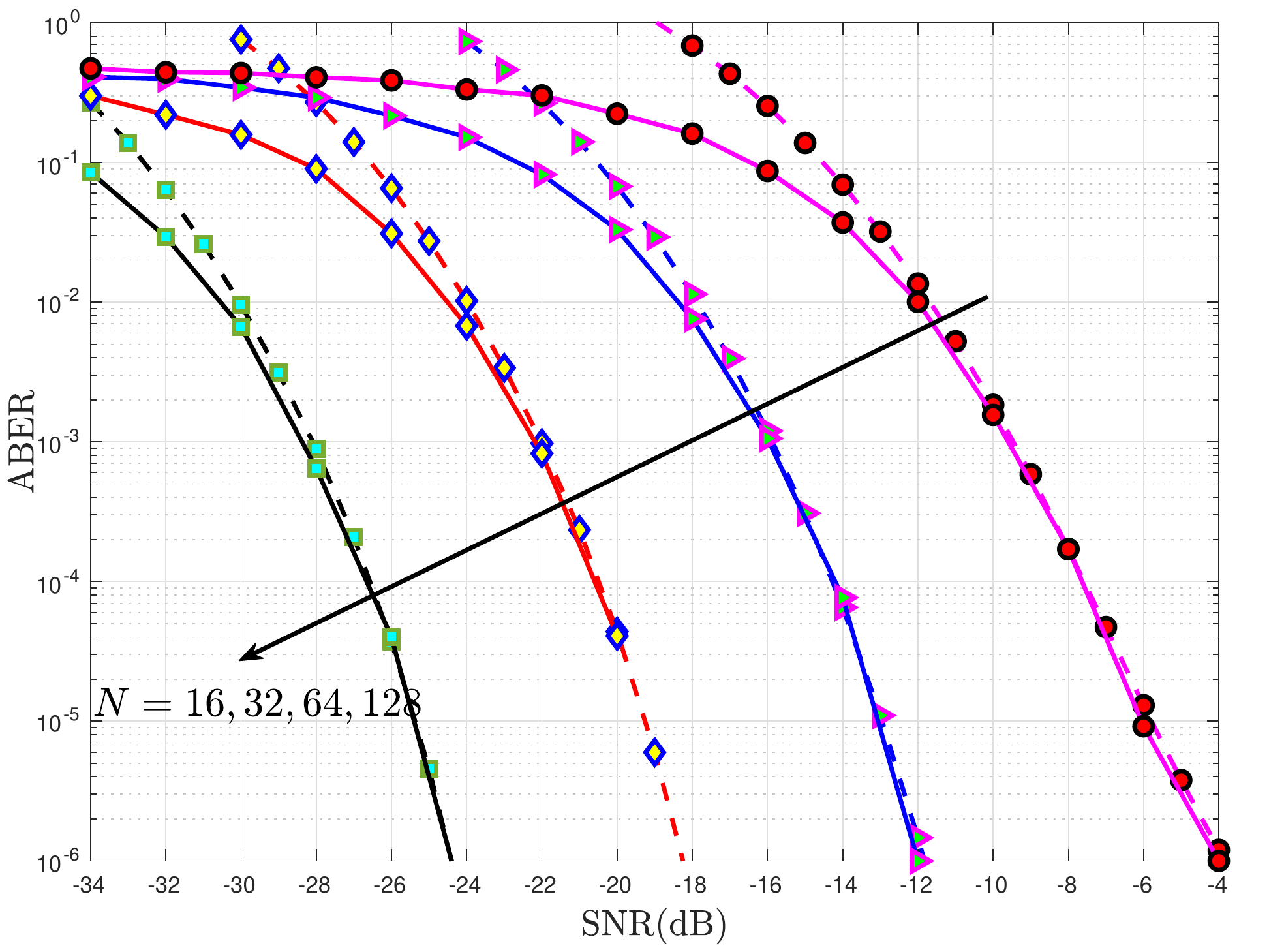}
        \caption{BER performance of the ASBC coded RIS based Received SM with $(M=4,N_r=2)$ for varying $N$.}
        \label{fig:ProposedSchemeBER2}
\end{figure}

We focus on the error performance of the proposed Alamouti Coded RIS-Based Received SM scheme in Fig.\ref{fig:ProposedSchemeBER} and Fig.\ref{fig:ProposedSchemeBER2}. These figures show the effect of the active reflecting elements number on the error performance for $M = 2$ and $M = 4$, respectively. The findings clearly show that The error performance linearly increases with the growth in N as expected. According to Fig.\ref{fig:ProposedSchemeBER}, the BER of proposed scheme significantly improves by about $18$ dB with the number of $N$ increasing from $16$ to $128$. Likewise, Fig.\ref{fig:ProposedSchemeBER} and Fig.\ref{fig:ProposedSchemeBER2} show that the effect of $M$ decreases with the increases $N$.

\begin{figure}
     \centering
\includegraphics[width=0.5\textwidth]{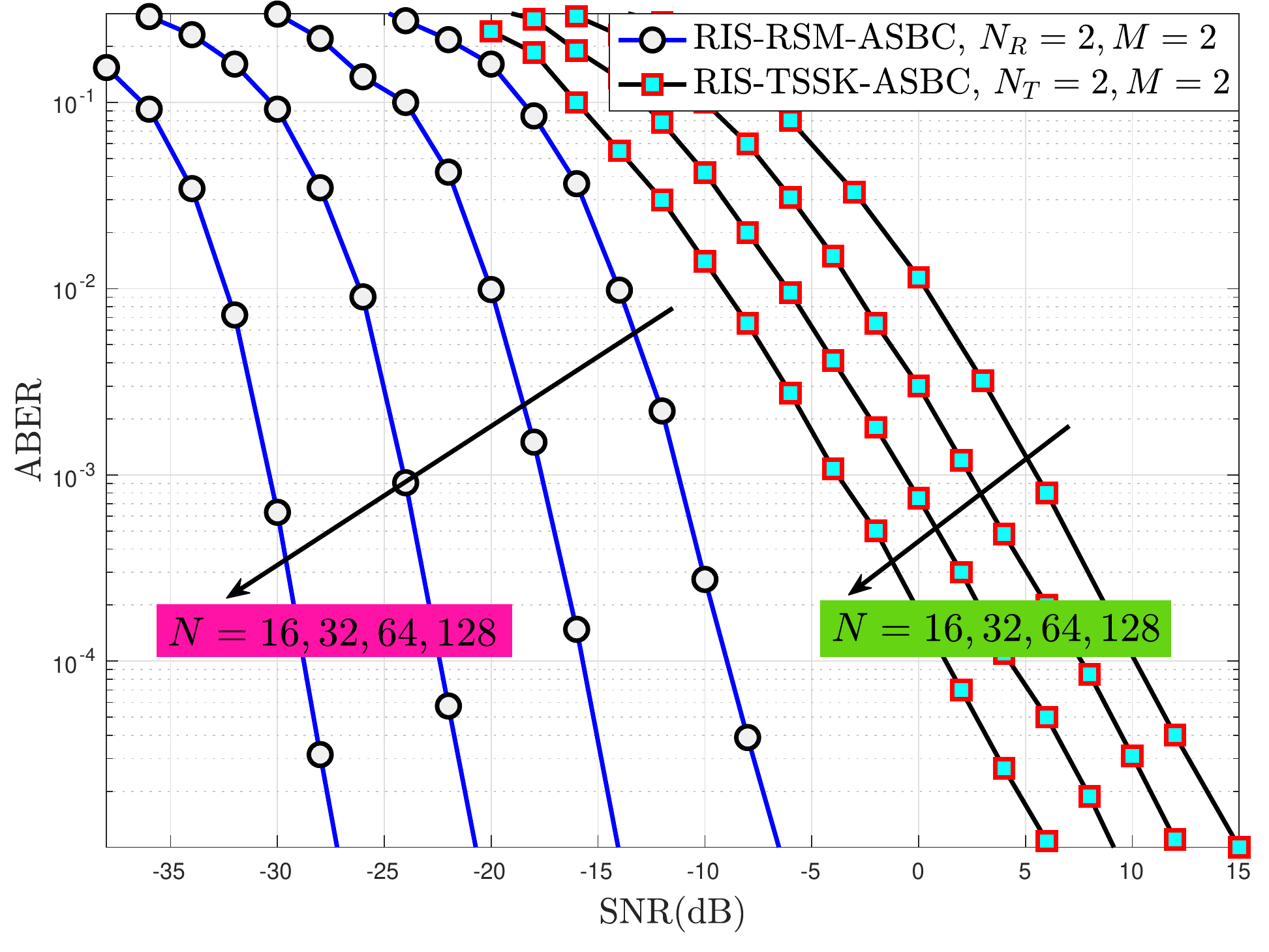}
        \caption{BER performance of the RIS-RSM-ASBC $(M=2,N_r=2 )$ and RIS-TSSK-ASBC $M=2,N_T=2,N_r=1)$ for varying $N$. }
        \label{fig:BERSSKRSSK}
\end{figure}

\begin{figure}
     \centering
\includegraphics[width=0.5\textwidth]{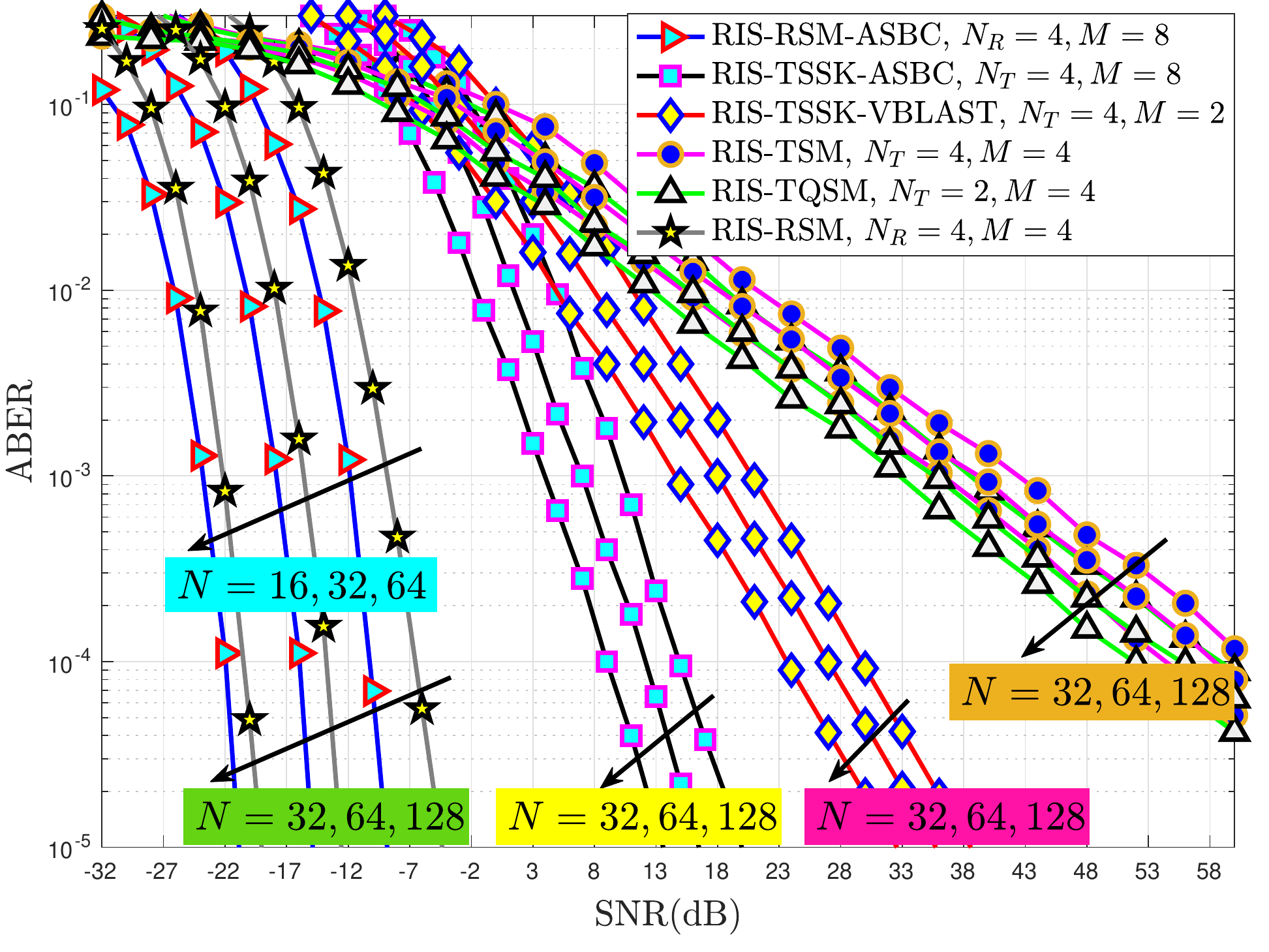}
        \caption{BER performance of the RIS-RSM ($M=8,N_r=4)$, RIS-SSK-ASTBC $( M=4,N_r=4)$, RIS-SSK-VBLAST $(M=8,N_T=4)$, RIS-SM $(M=2,N_T=4)$, RIS-QSM $(M=4,N_T=4)$ and  RIS-TQSM $(M=4,N_T=2)$ for varying $N$. }
        \label{fig:ComparedMult}
\end{figure}

As seen in these figures, there is a conformity between the simulation results and theoretical findings obtained from Fig. \ref{fig:ProposedSchemeBER} and Fig.\ref{fig:ProposedSchemeBER2}.

In Fig.\ref{fig:BERSSKRSSK}, we compare the BER performances of RIS-RSM-ASBC and RIS-TSSK-ASBC schemes. It is worth noting that the RIS-RSM-ASBC scheme provides about $14$ dB BER performance improvement over the RIS-TSSK-ASBC scheme. Also, we show the BER performance of the RIS-RSM-ASBC scheme at various $N$ in Fig.\ref{fig:BERSSKRSSK}. The error performances of both schemes improve with increasing $N$, while the effect of $N$ is more noticeable for the RIS-RSM-ASBC scheme. While increasing $N$ from $16$ to $128$ in the RIS-TSSK-ASBC scheme, the error performance improves about  $10$ dB, the BER performance of RIS-RSM-ASBC increases about $20$ dB. 

Fig.\ref{fig:ComparedMult} shows that RIS-RSM-ASBC provides better error performance contrasted to RIS-TSSK-ASBC, RIS-TSSK-VBLAST, RIS-TSM, RSI-TQSM, and RIS-RSM. It is worth noting that although there are fewer reflectors in the proposed scheme, it provides better performance than the conventional RIS-RSM system. As seen from Fig.\ref{fig:ComparedMult}, RIS-RSM-ASBC provides over $50$ dB improvement in the required SNR to achieve a target BER value in comparison to both RIS-TSM and RIS-TQSM. Also, it is shown that the proposed RIS-RSM-ASBC scheme is significantly better than RIS-TSSK-VBLAST. While RIS-RSM-ASBC provides $10^{-3}$ ABER performance at SNR $-17$ dB for $N = 32$, RIS-TSSK-VBLAST provides it with SNR $15$ dB for $N = 32$.

Lastly, we can summarize the performance improvements provided by the proposed scheme as follows:
\begin{itemize}
    \item RIS-RSM-ASBC scheme provides about $14$ dB BER performance improvements over RIS-TSSK-ASBC scheme.
    \item Compared to the proposed RIS-RSM-ASBC scheme, a more than $30$ dB difference in required SNR is observed for the RIS-TSSK-VBLAST.
    \item The proposed RIS-RSM-ASBC scheme provides over $50$ dB better performance than both RIS-TSM and RIS TQSM.
    \item We provide better error performance with the low number of reflectors in the proposed scheme in contrasted to the conventional RIS-RSM. 
\end{itemize}
\subsection{Complexity Analysis}
This section presents the receiver complexity of the specified/referenced systems and our proposed model. The computational detection complexity can be calculated by taking into account the real multiplication and summations of the receiver algorithms. The results of the complexity analysis are summarized in Table \ref{ComplexityAnalysis}. It is clear evident that classical Index Modulation systems have superior complexity performance than coded schemes.

\begin{table}
  \caption{ Complexity Analysis }
\centering
\begin{tabular}{l c c}
\rowcolor{cornflowerblue}
\hline\hline  
RIS-TSM && $(N+M)N^2_{t}$ Multiplication
    \\ \rowcolor{cornflowerblue}
      RIS-TQSM &&  $(N+M)N^2_{t}$ Multiplication
    \\ \rowcolor{cornflowerblue}
    RIS-RSM \cite{basar2020reconfigurable} && $(N+M)N^2_{r}$ Multiplication
    \\ \rowcolor{cornflowerblue}
    RIS-TSSK-ASBC \cite{li2021space} && $(N_tM)^2$ Multiplication
    \\  \rowcolor{cornflowerblue}
    RIS-TSSK-VBLAST \cite{li2021space} && $NN^2_{t}$ Multiplication
    \\\rowcolor{cornflowerblue}
  RIS-RSM-ASBC (Proposed System) && $(N/2+M)^{2}N^2_{r}$ Multiplication
    \\
    \hline
    \hline
\end{tabular}
\label{ComplexityAnalysis}
\end{table}

\section{Conclusions and Future Works}
In this paper, we have proposed Alamouti space-time coded RIS-assisted SM scheme at the receiver side and analytically derived the BER expressions of the proposed scheme. Also, we fully support the derived theoretical findings with Monte-Carlo simulations for different system parameters. Furthermore, we compared the error performance of the proposed scheme with RIS-TSM, RIS-TQSM, RIS-RSM, RIS-TSSK-ASBC and RIS-TSSK-VBLAST. According to our findings, RIS-RSM-ASBC significantly enhances the received SNR and provides ultra-reliable communication at extremely low SNR. Moreover, we have proposed the complexity analysis of RIS-RSM-ASBC.

As a future study, we want to examine the space-time block-coded RIS-based Received SSK and Received QSM schemes. Likewise, we can analyze the space-time block-coded RIS-based IM scheme with Deep Learning algorithm. Moreover, all these schemes can be analyzed with a sub-optimal Greedy Detector.

\bibliographystyle{IEEEtran}
\bibliography{IEEEabrv,references}

\end{document}